# Superconductivity in the Infinite Layer Compound $CaCuO_2$

K. A. Müller [1] and C. W. Chu [2]

[1] *Physik-Institut der Universität Zürich, CH-8057 Zürich, Switzerland*

[2] *Texas Center for Superconductivity-University of Houston, Houston. USA; Lawrence Berkeley National Laboratory, Berkeley, USA; and Hong Kong University of Science and Technology, Hong Kong, China.*

The infinite layer (IL) compound $CaCuO_2$ was first prepared in 1988.[1] It was thereafter investigated with interest regarding its superconducting properties. Recently, as a result of electric field-effect hole doping, a maximum $T_c$ of 89 K was reported by Schön *et al.*[2] The phase diagram this group obtained clearly resembles the one known in the cuprates upon chemical doping, especially YBCO. This similarity, when confirmed, may therefore establish electric field doping as a method to create high-temperature superconductivity as previously reported in organic layered materials[3] notably the fullerenes.[4]

In $CaCuO_2$ the $Cu^{2+}$ ions are fourfold coordinated by oxygens in the planes, and no apical oxygens are present. Schön *et al.* discuss their findings in terms of this IL property and the absence of apical oxygens above the Cu atoms. In doing so, they were not aware or ignored the photoemission work of Nücker *et al.*[5] This group investigated hole-doped $Sr_{0.67}Ca_{0.28}CuO_{2-\delta}$ by near-edge X-ray absorption spectroscopy (NEXAFS). By using polarized X-rays parallel or perpendicular to the $CuO_2$ planes, this technique can distinguish whether hole carriers are present on planar or on apical oxygens. The authors found that the characteristic peak for hole carriers at apical oxygens for E||c near 528 eV (Fig. 3. of Ref. 5), was absent when also superconductivity was absent. The latter was only observed by a small diamagnetic susceptibility signal in crystals, which disappeared after cutting off the surface layer. The authors concluded that the superconductivity observed in $CaCuO_2$ was due to the presence of apical oxygens in the surface layers of their crystals.

This is consistent with previous suggestions by Tao and Nissen that hole doped superconductivity in the IL compound is caused by inclusions of apical oxygen atoms.[6] In fact these two authors showed that a stacking sequence of $CuO_2$ / A / $CuO_2$ / $AO_\delta$ / Cu / $AO_\delta$ /$CuO_2$ / A / $CuO_2$ where the $AO_\delta$ represent a $\delta$ percentage of oxygens in a layer of A ($Sr^{2+}$ or $Ca^{2+}$) ions fitted much better existing previous HRTEM images than precedent simulations. The oxygen's in the $AO_\delta$ layer are in apical positions next to the $CuO_2$ layer. The above conclusions on the IL compound are supported by the earlier NEXAFS observations of Merz *et al.*[7] upon Ca doping of YBCO: In $Y_{1-x}Ca_xBa_2Cu_3O_y$ superconductivity and the 528 eV, E||c peak were absent simultaneously,

whereas upon oxygen doping both were present simultaneously.[7] The Karlsruhe group concluded that Ca doping of YBCO results in the absence of hole density at the apex oxygens and the absence of superconductivity, whereas oxygen doping results in the occurrence of hole density at the apex site and hence superconductivity.

With these results in mind, one has to ask whether in the experiments of Schön et al.[2] the last $CuO_2$ layer, and the one that according to the authors is responsible for the observed superconductivity, may not be „surface doped" with oxygen's. From Fig. 2 of Ref. 2 it is clear that the $CaCuO_2$ is covered by an $Al_2O_3$ layer. This highly insulating layer is necessary to apply the electric field to the $CaCuO_2$. For steric chemical reasons, it is very likely that oxygen's from the $Al_2O_3$ coverage are located above the $Cu^{2+}$ ions in the last $CuO_2$ layer, therefore forming fivefold-coordinated $Cu^{2+}$ ions <u>with apical oxygen's</u>. This likelihood is further supported because the observed maximum $T_c$ of 89 K is characteristic of hole-doped pyramidal coordinated $Cu^{2+}$, like in optimally doped YBCO.

In the electron-doped case of Schön et al.[2], the sample is the same and thus, by our reasoning, also the presence of apical oxygen. However it is well established from many experiments of electron-doped IL's, including the most recent ones [8], that apical oxygen is not present in that case.[9] Actually the maximum $T_c$ of 43 K in $Sr_{0.9}La_{0.1}CuO_2$ reported this year [8], and characteristic for these materials, is 9 K higher than the 34 K of Schön et al. This seems to indicate that the presence of apical oxygen is unfavorable for electron-doping as argued much earlier, because of the shorter *a* and/or *b* lattice-parameters.[10]

Whether the above suggestions can be verified remains to be seen. However, it is a question in a more general context, namely, that of which role the interface between the superconducting layer and the insulating $Al_2O_3$ layer plays, not only for $CaCuO_2$ but also in the electric field doped organic superconductors.[3,4] The latter materials are *per se* elastically more deformable, and thus it is possible that polaron formation occurs at the interface. Furthermore, is surface superconductivity, as proposed by de Gennes and observed for classical superconductors[11] involved, or discussed by Allender et al.[12] , but not yet detected in the semiconductor/ metal interface? In addition we recall that superconductivity at 90 K in $WO_3$ surface doped with Na was reported by Reich and Tsabba[13], and its existence confirmed by electron spin resonance (ESR) experiments.[14] A superconducting gap of $\Delta$ = 160 K was deduced by ESR, and later found from independent tunneling experiments to be 180K [15], i.e. within experimental error the same with a really different method.

One of us (KAM) acknowledges very interesting discussions with B. Batlogg, A. D. Bishop, and A. Shengelaya.